\documentclass[final,5p,times,twocolumn]{elsarticle}
\usepackage{amsmath}
\usepackage{amssymb} 

\usepackage{tikz}
\usetikzlibrary{arrows,matrix,positioning,patterns}
\usetikzlibrary{calc}
\usepackage{pgfplots} 
\usetikzlibrary{external}
\usetikzlibrary{angles}
\usepackage{hyperref}
\usetikzlibrary{plotmarks}

\newdefinition{rmk}{Remark}
\newproof{pf}{Proof}
\newproof{pot}{Proof of Theorem \ref{thm2}}

\usepackage[numbers]{natbib}
\usepackage[ruled, vlined, linesnumbered]{algorithm2e}

\usepackage{amsthm}

\newtheorem{problem}{Problem}

\usepackage{amsmath}
\usepackage{amssymb} 

\usepackage{eurosym}
\usepackage{colortbl}

\usetikzlibrary{arrows.meta,calc,decorations.markings,math,arrows.meta}

\usepackage[labelformat=simple]{subcaption}
\DeclareCaptionLabelSeparator{periodspace}{.\quad}
\usepackage{color,framed}
\usepackage[utf8]{inputenc} 
\usepackage{tabularx}

\usepackage{rotating}
\usepackage{graphicx}
\usepackage{lscape}
\allowdisplaybreaks 
\usepackage{longtable}

\usepackage[utf8]{inputenc}
\usepackage{pgfplots}
\usepgfplotslibrary{groupplots} 
\pgfplotsset{compat=1.3}

\makeatletter
\def\ps@pprintTitle{%
 \let\@oddhead\@empty
 \let\@evenhead\@empty
 \def\@oddfoot{}%
 \let\@evenfoot\@oddfoot}
\makeatother

\begin{document}

\begin{frontmatter}
%
%
\title{K-subareas partitioning subject to coinciding boundary segments of neighboring subareas}
\author{Mogens Plessen\corref{cor1}}
\cortext[cor1]{MP is with Findklein GmbH, Switzerland, \texttt{pmogens@proton.me}}

%
%
%
%

\begin{abstract}
A method for the problem of partitioning an area into K subareas subject to coinciding boundary segments of neighboring subareas is presented. The given area may be non-convex and may additionally include obstacle areas. The proposed hierarchical method involves three steps: repeated bisection splitting, merging and the solution of traveling salesman problems. Geographic real-world outdoors areas data is used for the evaluation of the method. The benefit of computational simplicity at the cost of imbalanced subarea sizes is highlighted.
\end{abstract}
\begin{keyword}
Area partitioning; Geographic information systems; Area coverage; Agriculture.
\end{keyword}
\end{frontmatter}


\section{Introduction\label{sec_intro}}

\begin{table}
\centering
\begin{tabular}{|ll|}
\hline
\multicolumn{2}{|c|}{MAIN NOMENCLATURE}\\
\multicolumn{2}{|l|}{Symbols}\\
$K$ & Desired number of subareas, (-).\\
$\hat{K}$ & Number of subareas after the splitting step, (-)\\
$N_p$ & Power coefficient, (-).\\[3pt]
\multicolumn{2}{|l|}{Abbreviations}\\
TSP & Traveling Salesman Problem.\\
\hline
\end{tabular}
\end{table}

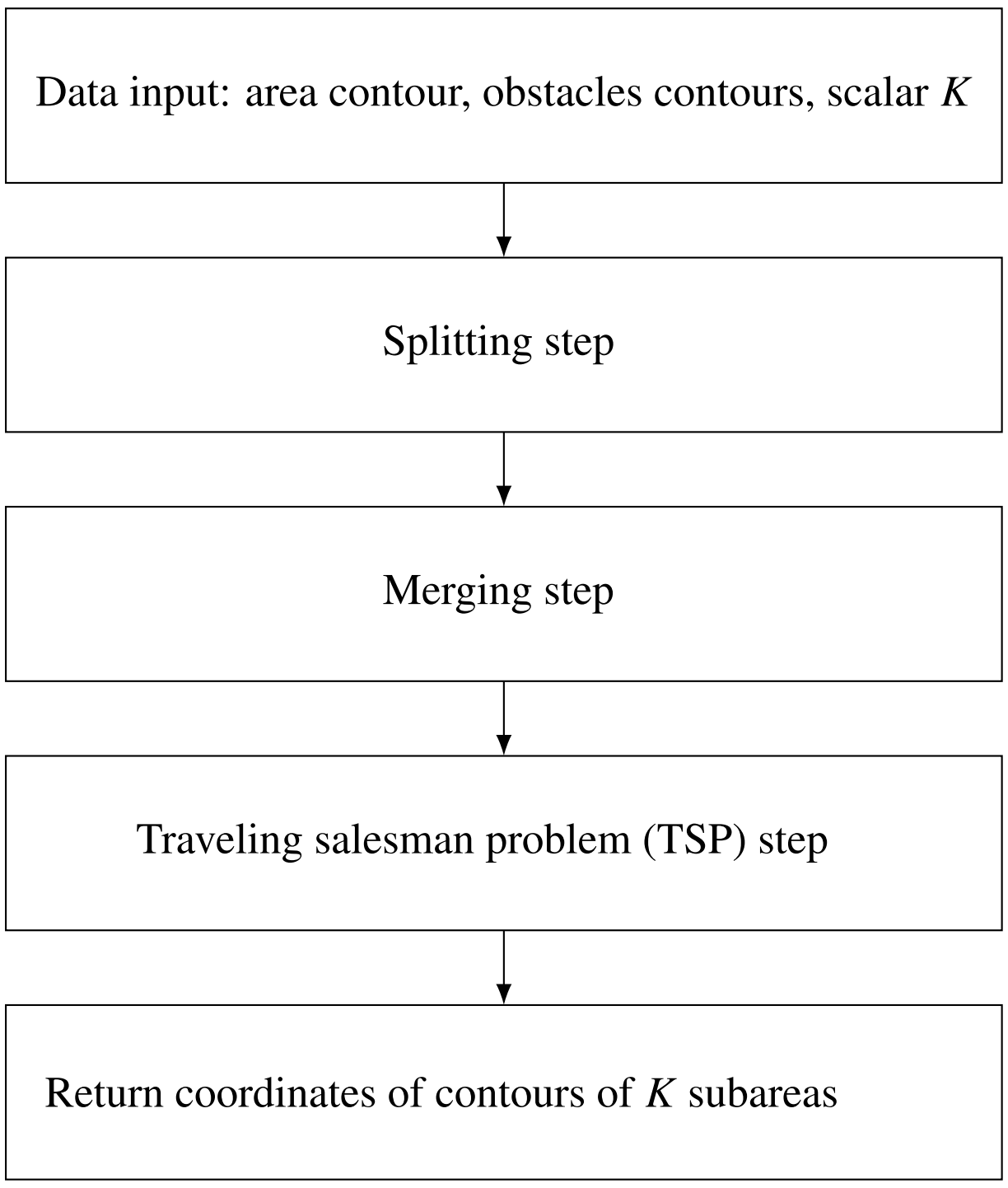
\begin{figure}
\vspace{0.3cm}
\centering%
\begin{tikzpicture}
\draw[fill=white] (0,8) rectangle (8,9.4);
\node (c) at (4,8.7) [scale=1,color=black,align=left] {Data input: area contour, obstacles contours, scalar $K$};
\draw[black,-{Latex[scale=1.0]}] (4, 8) -- (4, 7.4);
\draw[fill=white] (0,6) rectangle (8,7.4);
\node (c) at (4,6.7) [scale=1,color=black,align=left] {
Splitting step
};
\draw[black,-{Latex[scale=1.0]}] (4, 6) -- (4, 5.4);
\draw[fill=white] (0,4) rectangle (8,5.4);
\node (c) at (4,4.7) [scale=1,color=black,align=left] {
 Merging step
};
\draw[black,-{Latex[scale=1.0]}] (4, 4) -- (4, 3.4);
\draw[fill=white] (0,2) rectangle (8,3.4);
\node (c) at (3.87,2.7) [scale=1,color=black,align=left] {
Traveling salesman problem (TSP) step
};
%
\draw[black,-{Latex[scale=1.0]}] (4, 2) -- (4, 1.4);
\draw[fill=white] (0,0) rectangle (8,1.4);
\node (c) at (3.5,0.7) [scale=1,color=black,align=left] {Return coordinates of contours of $K$ subareas};
\end{tikzpicture}
\caption{High-level algorithm of this paper as a block diagram. Data input comprises an area contour, potential obstacles areas contours coordinates, as well as a desired number of subareas $K$ after area partitioning.}
\label{fig_blockdiag}
\end{figure}

The technique of partitioning a given area into multiple subareas can have many applications. For example for distributed area coverage applications, a work area may be partitioned, before each subarea is covered by a separate robot simultaneously.

For coverage path planning, a popular partitioning method is the Boustrophedon cellular decomposition technique \cite{choset1998coverage}. However, the resulting number of partitioned subareas cannot be prescribed and is instead determined by obstacles in the work space, which create separate cells after a line is swept over the work space. Thus, (i) a desired number of subareas, and (ii) desired subarea sizes (e.g., roughly equal) cannot be prescribed, but result naturally from the line sweep and the specific work space characteristics. 

\begin{figure}
\centering
\includegraphics[width=.99\linewidth]{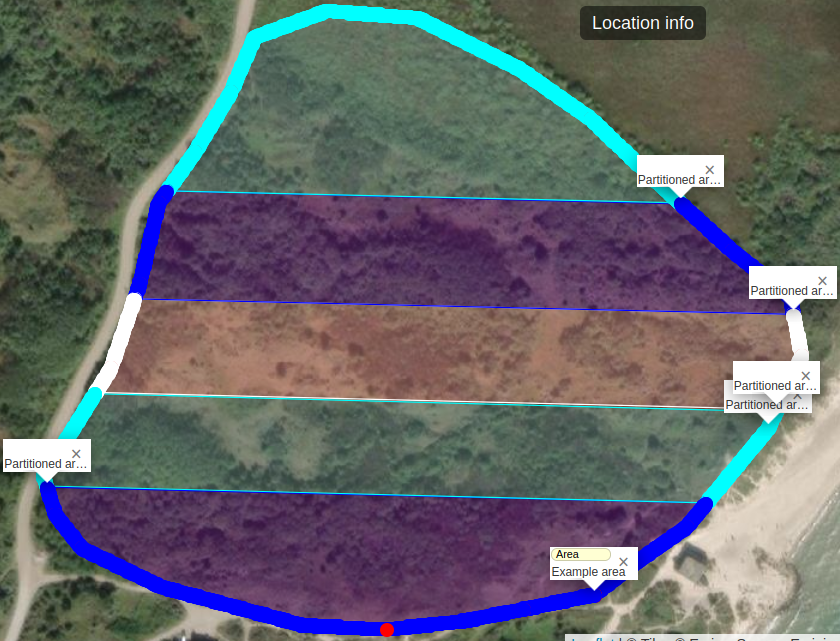}
\caption{Illustration of an alternative method: a sweep-line algorithm. The method (i) assigns desired subarea sizes (e.g. equal sizes), before (ii) sweeping a line over the area and partitioning subareas as soon as desired subarea sizes are attained. This method works optimally for convex areas, but fails for non-convex areas where a single line sweep is in general insufficient since it may generate disconnected areas for a given line sweep section.}
\label{fig_ex1_alternative}
\end{figure}

For convex areas, a sweep-line approach permits to optimally partition a given area into multiple subareas; see Fig. \ref{fig_ex1_alternative} for illustration. However, for the partitioning of non-convex areas a single swept line in general fails since it may generate disconnected areas for a given sweep-line section.

\begin{figure}
\centering
\begin{subfigure}[t]{.24\textwidth}
  \centering
  \includegraphics[width=.99\linewidth]{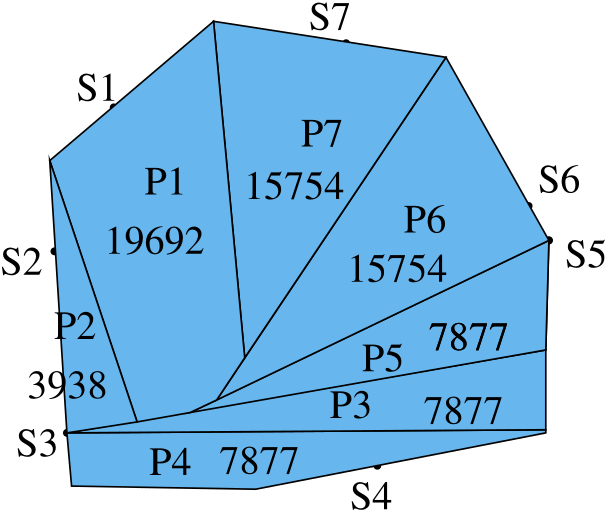}
\caption{Visualization of triangular-shaped subareas exhibiting very steep wedges. Such subareas are in general undesired since difficult to cover with nonholonomic vehicles with, for example, limited turning radii.}
  \label{fig_counterex1}
\end{subfigure}~~~%
\begin{subfigure}[t]{.24\textwidth}
  \centering
  \includegraphics[width=.99\linewidth]{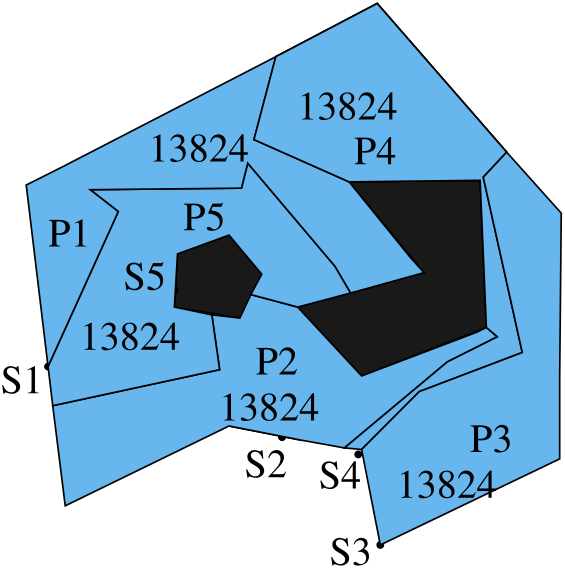}
\caption{Visualization of highly non-convex subareas. Note in particular subarea P4, which continues just up to site S4, and which becomes partially extremely narrow.}
  \label{fig_counterex1}
\end{subfigure}%
\caption{Illustration of an alternative method: examples from \cite{hert1998polygon}. Two disadvantages are visualized.}
\label{fig_counterex}
\end{figure}

Extending to the non-convex area domain, advanced sweep-line based algorithms are a popular approach for area partitioning \cite{hert1998polygon, bast2000area, huang2001optimal, xing2020area, skorobogatov2021multi, abrahamsen2025partitioning, huck2026qgis}. However, and similarly to triangular subareas used in \cite{kang2025multi}, two observations were made. First, polygon areas considered were often very sparse and consisting only of very few vertices along the area contour, which resulted in boundary segments of very edgy shape. Second, calculated subareas often exhibited very steep wedges or triangular-shaped subareas, which may be difficult to cover with nonholonomic vehicles. See Fig. \ref{fig_counterex} for visualization.


Alternatively, in \cite{baraniak2026efficient} a gradient-based rectangular area decomposition method is presented to segment the map into rectangular zones. In \cite{li2023full} partitioning into convex polygons is considered. A discretization grid-based approach is presented in \cite{santamaria2013rapid}, where a starting cell is selected and then grown by selecting neighboring cells. The difficulty of such an approach is how to select the starting cells and how to avoid isolated cells, that may result at the boundaries of a partitioned area when it is reaching its designated size. Other grid-based approachs are \cite{skorobogatov2025compact} and \cite{wzorek2021polygon}, whereby the latter additionally proposes a compactness metric.

Above literature review revealed a reasearch gap with the absence of a clustering-based approach in combination with the additional constraint of maintaining coinciding boundary segments between neighboring subareas. This observation provided the motivation for this article.

In this article a heuristic and hierarchical simple method is presented for the area partitioning problem. The following two additional assumptions are made: (i) The given area may be non-convex and may include obstacle areas. (ii) Boundary segments of neighboring subareas must be coinciding.

The remaining paper is organised as follows: problem formulation, modeling and proposed solution, discussion and the conclusion are described in Sections \ref{sec_materials}-\ref{sec_conclusion}.

\section{Materials and methods\label{sec_materials}}

\subsection{Problem formulation\label{subsec_probformulation}}

The problem addressed can be formulated as follows.

\begin{problem}\label{problem1}
Given an area that may in general be non-convexly shaped and may include obstacle areas, calculate a partitioning of the area into $K$ subareas of roughly equal sizes subject to coinciding boundary segments of neighboring subareas.
\end{problem}

\subsection{Discussion of constraints\label{subsec_cstrts}}

The main difficulty of the problem stems from the constraint seeking coinciding boundary segments of neighboring subareas in combination with the assumption of non-convex given areas.

To see this, consider for contrast a discretized area partitioning that labels coordinates, for example resulting from a sampled meshgrid, in the interior of the area contour.  Then, computing the hulls of the group of coordinates with the same label would generate subarea contours, but these are not coinciding for boundary segments of neighboring subareas.

However, coinciding boundary segments of neighboring subareas are crucial for area coverage applications because only then the distributed coverage of all subareas would permit full area coverage of the original area without area coverage gaps.

\subsection{Proposed solution\label{sec_solnMethod}}


\begin{figure*}
\centering
\begin{subfigure}[t]{.33\textwidth}
  \centering
  \includegraphics[width=.99\linewidth]{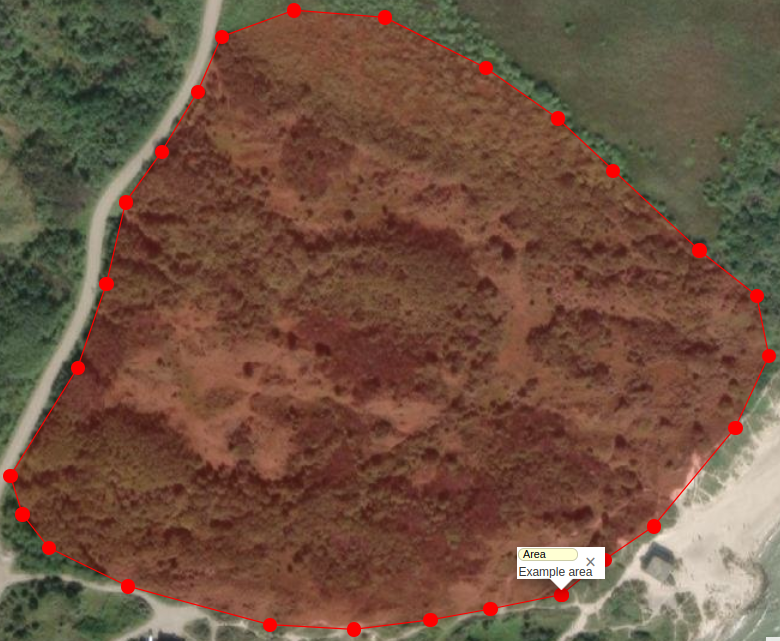}
\caption{Input: area contour data, scalar $K=3$.}
  \label{fig_explain_1}
\end{subfigure}%
\begin{subfigure}[t]{.33\textwidth}
  \centering
  \includegraphics[width=.99\linewidth]{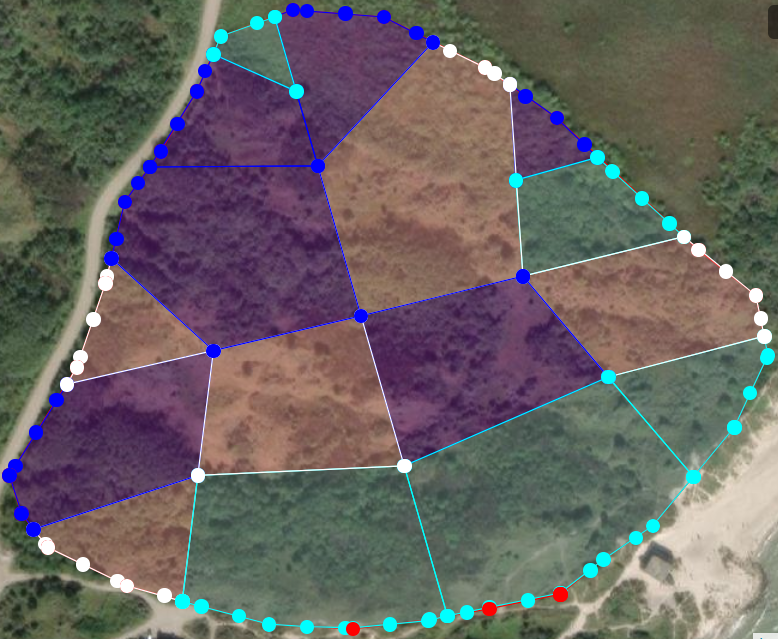}
\caption{Result after the splitting step for $N_p=4$.}
  \label{fig_explain_2}
\end{subfigure}%
\begin{subfigure}[t]{.33\textwidth}
  \centering
  \includegraphics[width=.99\linewidth]{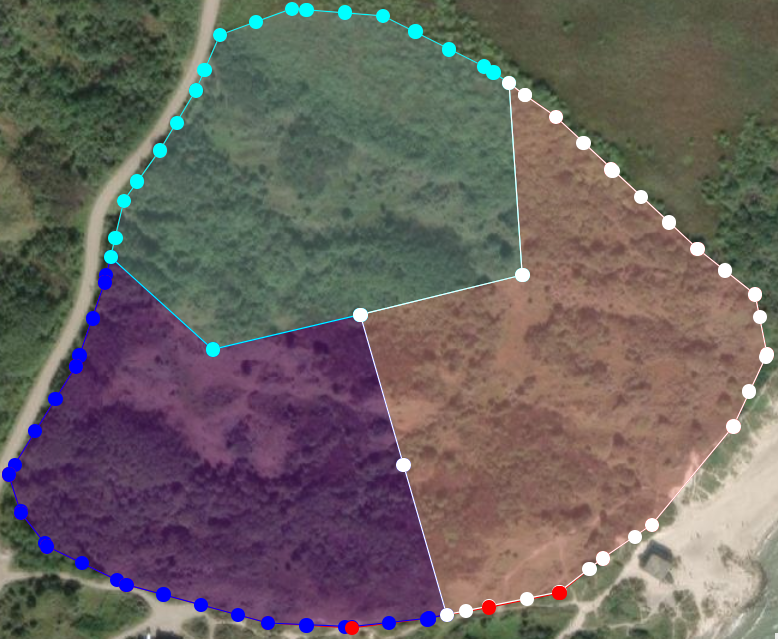}
\caption{Result after the merging and TSP-steps.}
  \label{fig_explain_3}
\end{subfigure}
\caption{Exemplary illustration of the main steps of the  algorithm according to Fig. \ref{fig_blockdiag}. A given area is partitioned into $K=3$ subareas.}
\label{fig_explain}
\end{figure*}

%


\begin{figure*}
\centering
\begin{subfigure}[t]{.33\textwidth}
  \centering
  \includegraphics[width=.99\linewidth]{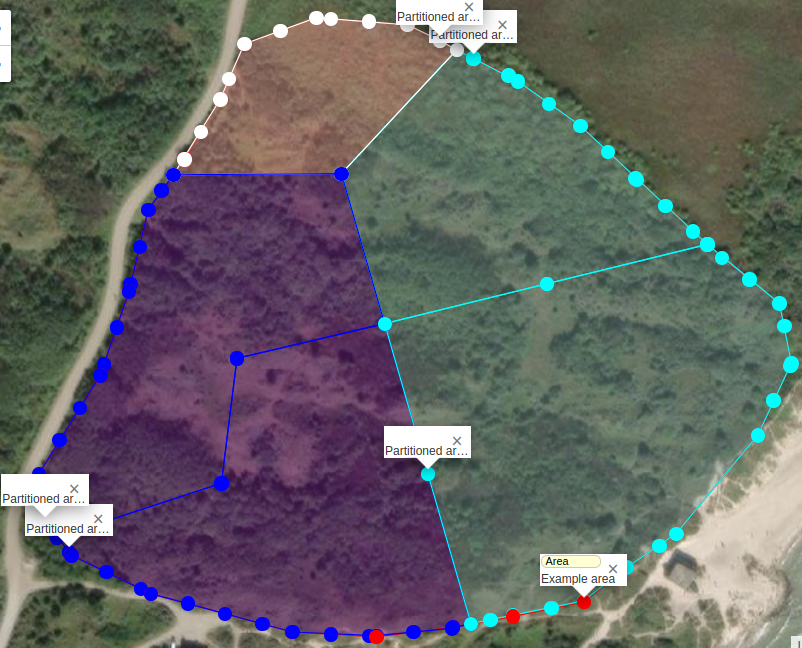}
\caption{$N_p=4$.}
  \label{fig_Ex1_power4}
\end{subfigure}%
\begin{subfigure}[t]{.33\textwidth}
  \centering
  \includegraphics[width=.99\linewidth]{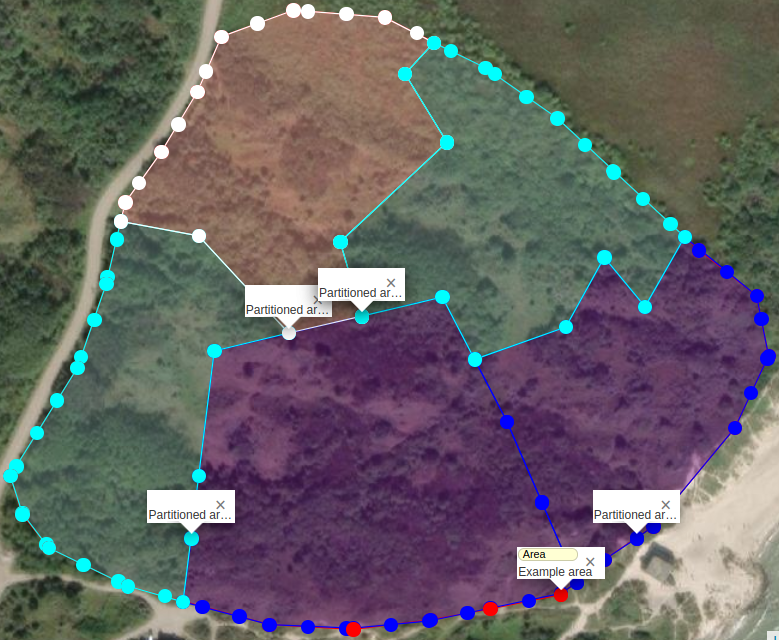}
\caption{$N_p=6$.}
  \label{fig_Ex1_power6}
\end{subfigure}%
\begin{subfigure}[t]{.33\textwidth}
  \centering
  \includegraphics[width=.99\linewidth]{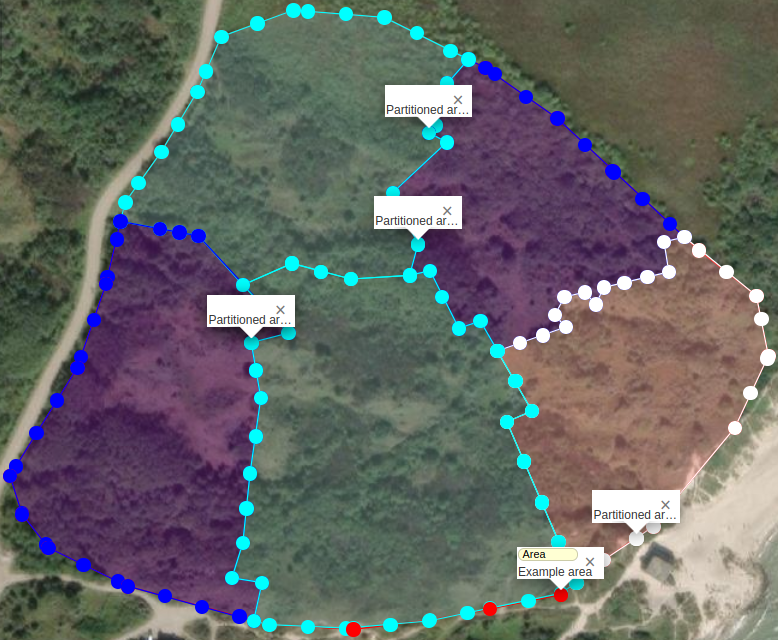}
\caption{$N_p=9$.}
  \label{fig_Ex1_power9}
\end{subfigure}
\caption{Example 1. Illustration of the influence of hyperparameter $N_p$ on solution quality.}
\label{Ex1_power469}
\end{figure*}

%

The proposed high-level algorithm is summarised as a block diagram in Fig \ref{fig_blockdiag} and further visualised in Fig. \ref{fig_explain}. It is composed of three steps: splitting, merging and the solution of traveling salesman problems. These are presented in the following. A discussion is provided in the subsequent Sect. \ref{sec_discussion}.

First, starting from the initial area the splitting step implies repeated bisection into subareas until 
\begin{equation}
\hat{K}=2^{N_p}\label{eq_hatK}
\end{equation}
different subareas are obtained. Thus, repeated bisection here implies geometrically splitting an area into two equal regions at each step. This is obtained by (i) optionally adding to the points of the area contour additional auxiliary points within the valid polygonal domain of each area (and avoiding the interior of obstacle areas) to improve the spatial distribution of points, before (ii) clustering the resulting points into two groups, and (iii) tracing the contours of these regions and its partitioning and coinciding boundary line to obtain the two bisected subareas. Note that integer $N_p$ is a hyperparameter. It can either be assigned as a constant, for example $N_p=6$, or made be variable as a function of input $K$. For example, a heuristic may be to increase integer $N_p$ as long a $2^{N_p}<7K$, where 7 is a scaling hyperparameter. $N_p$ is the hyperparameter dominating the computational complexity of the algorithm according to Fig. \ref{fig_blockdiag}. This will be illustrated in a numerical example later on. Note that the key reason why repeated bisection is proposed, is that because any bisection permits to easily maintain track of coinciding subarea boundary segments. This is since for a bisected area the coinciding boundary of the resulting two subareas is just a line segment.

Second, for $K=\hat{K}$ no further work is required and the result can be returned, however, for arbitrary $K$ a merging step follows. It is composed of (i) clustering the centroids of the $\hat{K}$ regions according to \eqref{eq_hatK} into $K$ groups, before (ii) applying a filtering step to maintain only area contour coordinates and area coordinates of obstacle contours that cut the area contour. For clustering the classic $K$-means algorithm is used. As heuristic implementation details and introducing hyperparameters, the initial $K$ means are initialized randomly as coordinates among the $\hat{K}$ centroids, the algorithm is restarted 100 times, and a maximum iteration limit of 100 is set in case a restart iteration did not converge before. 

Third, a traveling salesman problem (TSP) step is applied. Therefore, a TSP is solved for each of the $K$ subareas to order the contour coordinates. The implementation follows to method from \cite{plessen2025path}.

Finally, the algorithm according to Fig. \ref{fig_blockdiag} returns the coordinates of the the contours of the $K$ subareas.

%

\section{Discussion and limitations\label{sec_discussion}}

In this section results of numerical experiments for real-world data are visualized to highlight features of proposed algorithm and discuss its limitations.

\subsection{Numerical results for real-world data}


\begin{figure}
\centering
\begin{subfigure}[t]{.25\textwidth}
  \centering
  \includegraphics[width=.99\linewidth]{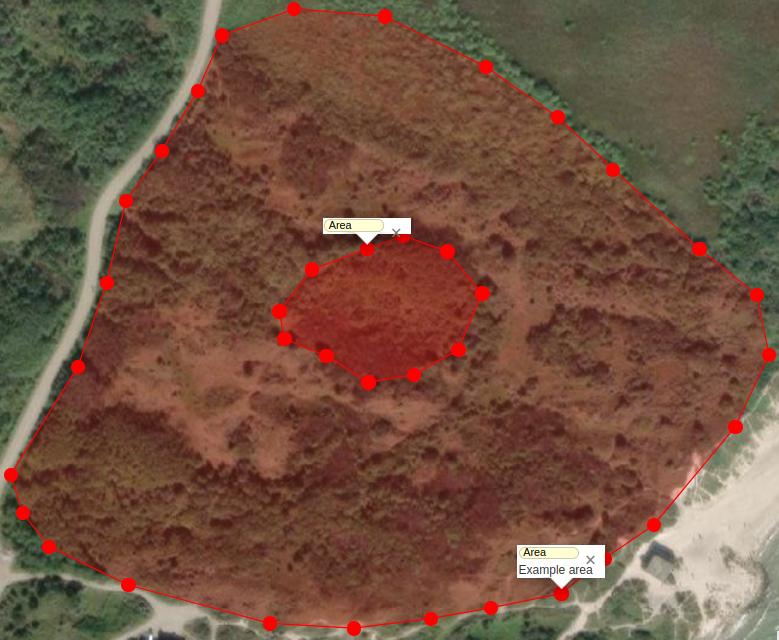}
\caption{Area and obstacle contour data.}
  \label{fig_Ex1_power4}
\end{subfigure}%
\begin{subfigure}[t]{.25\textwidth}
  \centering
  \includegraphics[width=.99\linewidth]{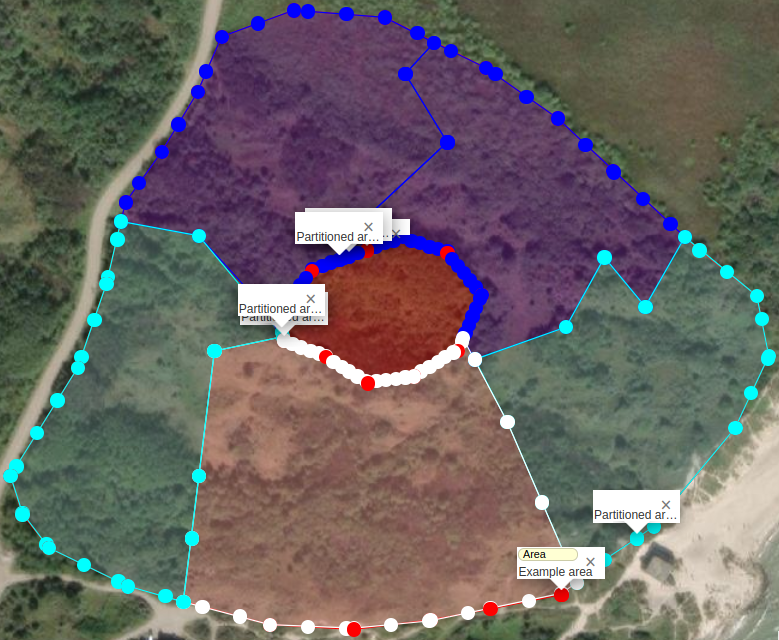}
\caption{Result for $K=5$.}
  \label{fig_boustro_noheadl}
\end{subfigure}%
\caption{Example 2. Illustration of obstacle area handling.}
\label{Ex2}
\end{figure}
%

%
\begin{figure}
\centering
\begin{subfigure}[t]{.25\textwidth}
  \centering
  \includegraphics[width=.99\linewidth]{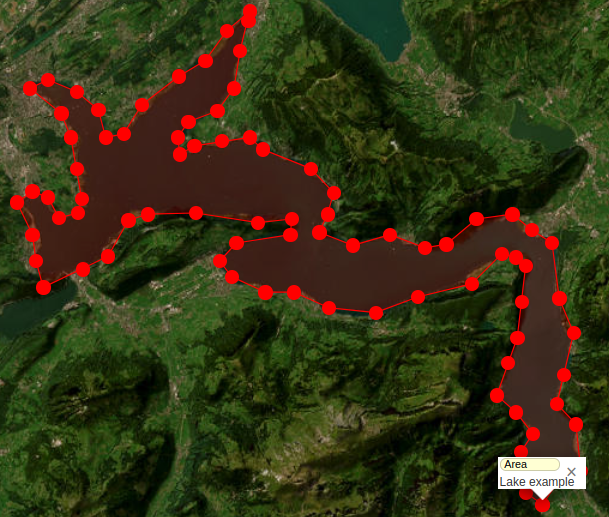}
\caption{Area contour data.}
  \label{fig_Ex1_power4}
\end{subfigure}%
\begin{subfigure}[t]{.25\textwidth}
  \centering
  \includegraphics[width=.99\linewidth]{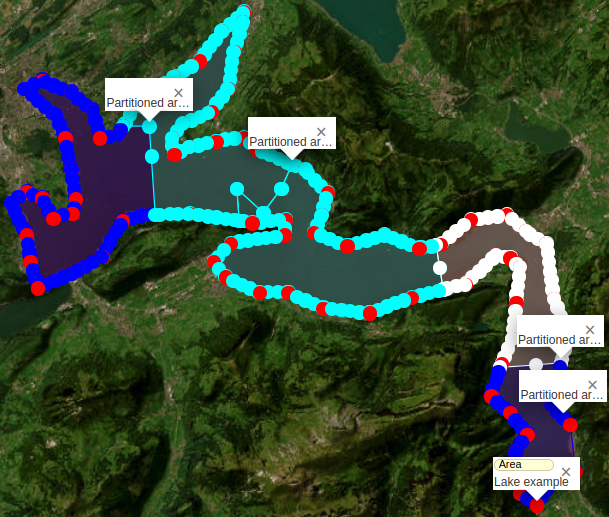}
\caption{Result for $K=5$.}
  \label{fig_boustro_noheadl}
\end{subfigure}%
\caption{Example 3. Highly non-convex area contour of a lake.}
\label{Ex3_lake_2figs}
\end{figure}
%


\begin{figure}
\centering
\begin{subfigure}[t]{.25\textwidth}
  \centering
  \includegraphics[width=.99\linewidth]{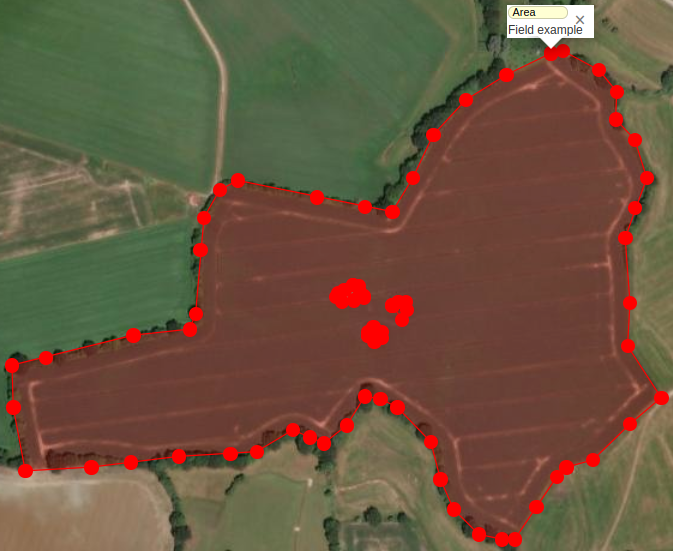}
\caption{Input data.}
  \label{fig_Ex4_data}
\end{subfigure}%
\begin{subfigure}[t]{.25\textwidth}
  \centering
  \includegraphics[width=.99\linewidth]{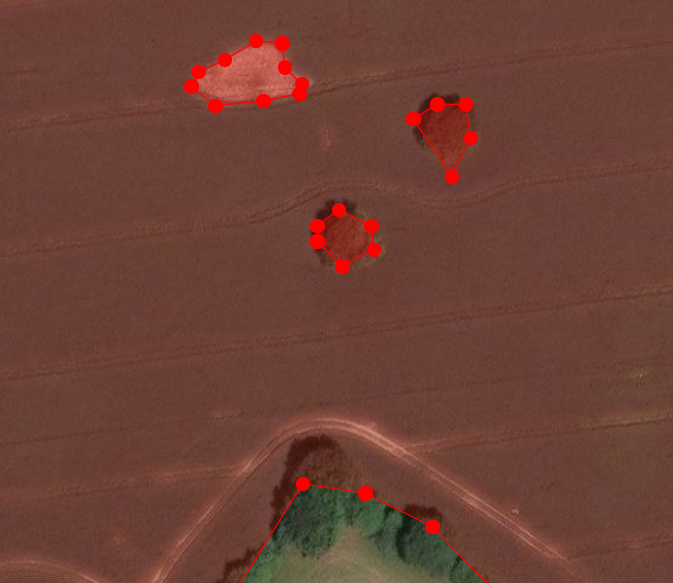}
\caption{Zoom-in to obstacle areas.}
  \label{fig_Ex4_obstacles_zoom}
\end{subfigure}%
\caption{Example 4. Illustration of input data composed of an agricultural field contour and 3 obstacle area contours.}
\label{Ex4_data}
\end{figure}

\begin{figure}
\centering
\begin{subfigure}[t]{.25\textwidth}
  \centering
  \includegraphics[width=.99\linewidth]{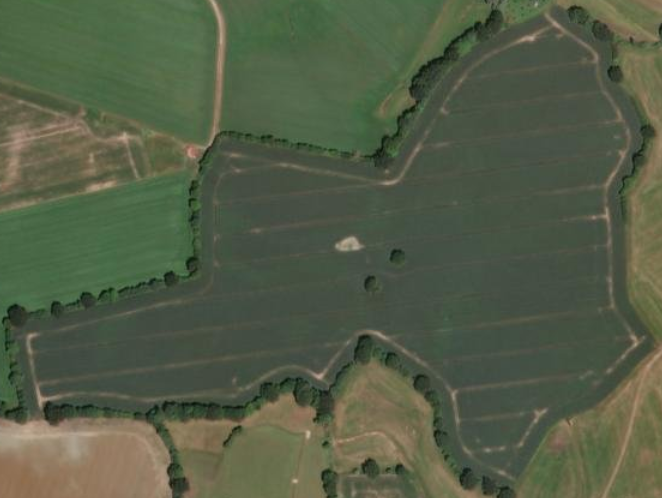}
\caption{A field with 3 obstacle areas.}
  \label{fig_Ex4_pure}
\end{subfigure}%
\begin{subfigure}[t]{.25\textwidth}
  \centering
  \includegraphics[width=.99\linewidth]{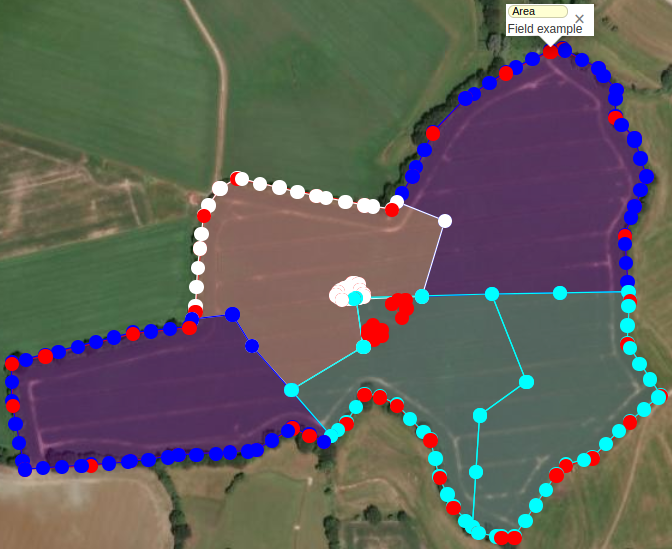}
\caption{Result for $K=5$.}
  \label{fig_Ex4}
\end{subfigure}%
\caption{Example 4. Illustration of obstacle area handling by the algorithm. In this example, 2 obstacle areas (red) are included in the interior of one of the $K$ subareas, and 1 obstacle area splits two subareas. As a result, a segment of its boundary becomes part of the boundary of one subarea (white), and another segment of its boundary becomes part of the boundary of another subarea (cyan).}
\label{Ex4_results}
\end{figure}

%


\begin{figure}
\centering
\begin{subfigure}[t]{.25\textwidth}
  \centering
  \includegraphics[width=.99\linewidth]{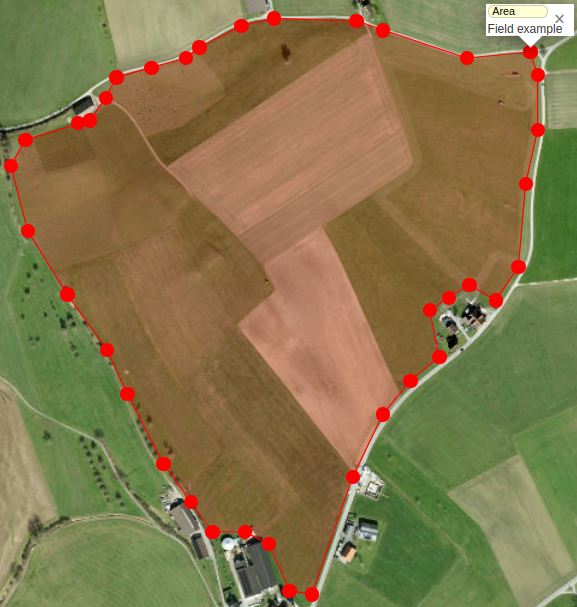}
\caption{Input data.}
  \label{fig_Ex1_power4}
\end{subfigure}%
\begin{subfigure}[t]{.25\textwidth}
  \centering
  \includegraphics[width=.99\linewidth]{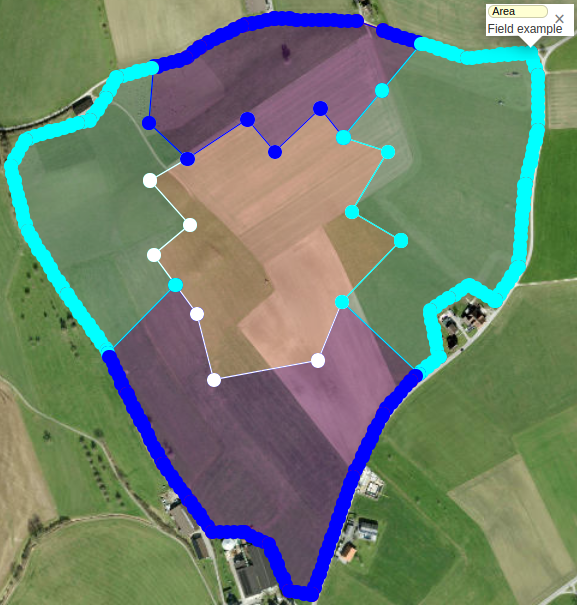}
\caption{Result for $K=5$.}
  \label{fig_boustro_noheadl}
\end{subfigure}%
\caption{Example 5. Illustration of an application where an agricultural work area composed of multiple existing fields is partitioned differently.}
\label{Ex5}
\end{figure}
%
%



\begin{figure}
\centering
\begin{subfigure}[t]{.25\textwidth}
  \centering
  \includegraphics[width=.99\linewidth]{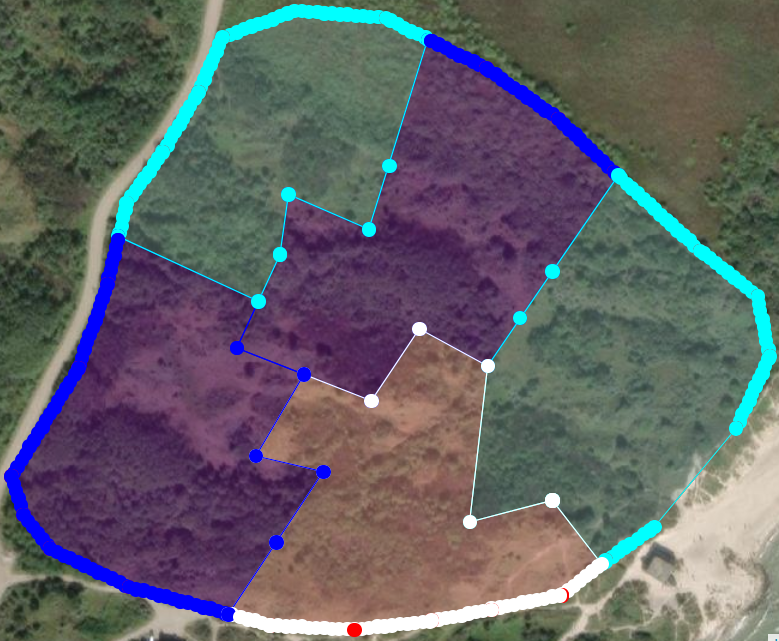}
\caption{$N_p=6$: $e_\text{max}=18.4\%,~\tau=1.5$s.}
  \label{fig_Ex6_K6_davgdiv10}
\end{subfigure}%
\begin{subfigure}[t]{.25\textwidth}
  \centering
  \includegraphics[width=.99\linewidth]{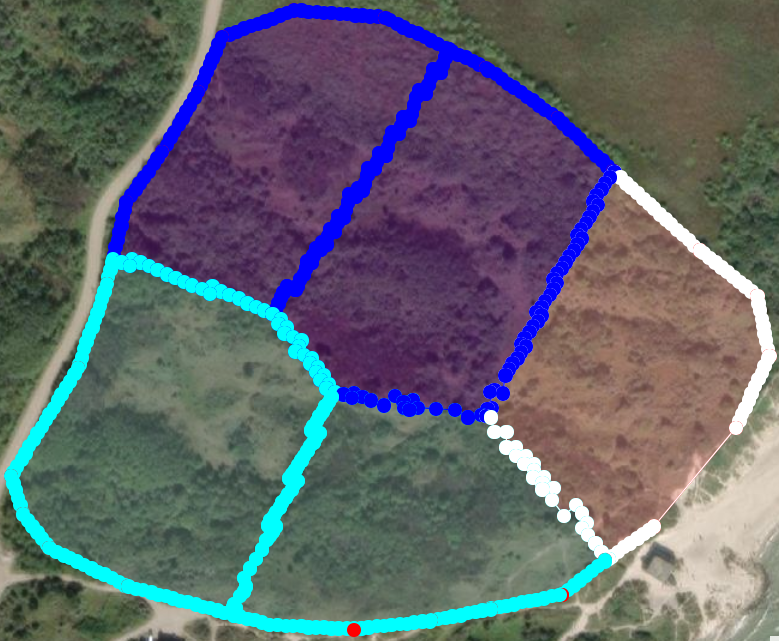}
\caption{$N_p=13$: $e_\text{max}=12.1\%,~\tau=217.9$s.}
  \label{fig_Ex6_K13_davgdiv10}
\end{subfigure}%
\caption{Example 6. Illustration of the effects of large $N_p$ influencing (i) visual subarea contour jaggedness, (ii) computational runtime complexity $\tau$, and (iii) area partitioning quality with $e_\text{max}$ denoting the largest relative error over all subareas with respect to the ideal partitioned subarea size. A trade-off between more balanced subarea sizes and computational complexity is observed.}
\label{Ex6}
\end{figure}
%
%

%

\begin{figure}
\centering
\includegraphics[width=.99\linewidth]{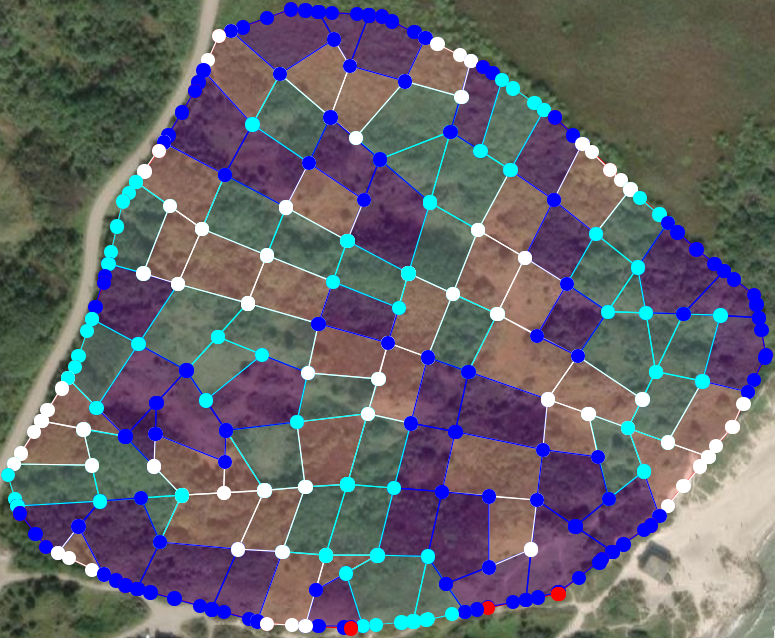}
\caption{Example 7. Large-scale example with partitioning into $K=100$ subareas.}
\label{Ex7}
\end{figure}

Numerical results for real-world data are displayed in Fig. \ref{Ex1_power469} - \ref{Ex7}. Fig. \ref{Ex1_power469} illustrates how increasing $N_p$ refines solution quality. Fig. \ref{Ex2} illustrates how an obstacle area splits subareas and generates new boundary segments. Fig. \ref{Ex3_lake_2figs} illustrates subarea partitioning of a highly non-convex area. One important insight for these type of examples was the importance of a densely discretized area contour to help the TSP-step trace subarea contours. Respective limitations of the algorithm according to Fig. \ref{fig_blockdiag} will be further discussed in the subsequent section \ref{subsec_limitations}. Fig. \ref{Ex4_data} and \ref{Ex4_results} illustrate results for an example with three obstacle areas. Fig. \ref{Ex5} illustrates an application, where an existing agricultural work area may be partitioned differently. Fig. \ref{Ex6} illustrates computational complexity and balancing quality. Therefore, the following two properties are defined,
\begin{equation}
e_\text{max}=\max_{k\in\{1,\dots,K\}} |\frac{A_k - \bar{A}}{\bar{A}} |,
\end{equation}
where $\bar{A}=A/K$, and $\tau$ denoting the computational runtime required for the algorithm according to Fig. \ref{fig_blockdiag}. Experiments were run on a laptop equipped with an Intel Core i9 CPU @5.50 GHz×32 and 32 GB of memory.

\subsection{Dicsussion of hyperparameters and limitations\label{subsec_limitations}}

It was found that the performance of the algorithm according to Fig. \ref{fig_blockdiag} is largely determined by two main hyperparameters. These are $N_p$ and the inter-point distance of area contours input data. Larger $N_p$ improve more balanced subarea sizes but drastically increase computationally complexity; cf. Fig. \ref{Ex6}. Smaller inter-point spacing, which can be aided by interpolating existing data, consistently improved performance but increases the overall data size.

The option of adding additional auxiliary points within the polygonal domain of each region to improve the spatial distribution of points before clustering was found to have only marginal influence on results. Throughout all experiments a grid spacing of $\Delta d=\text{min}(\Delta_x^\text{max},\Delta_y^\text{max})/5$ is used, where $\Delta_x^\text{max}$ and $\Delta_y^\text{max}$ denote the domain range of the input area contour along the $x$- and $y$-axis, respectively.

The proposed algorithm is limited by the implementation of the method to solve TSPs, which are NP-hard problems. Therefore, while theoretically excluded, in practice and  depending on the TSP-solution algorithm implementation and depending on given input data, TSP-tours may in general be found that result in boundary segments that overlap for neighboring subareas.

\subsection{Outlook\label{subsec_outlook}}

As Fig. \ref{Ex6} highlighted, there is a tradeoff between low computational complexity and balanced subarea sizes and less edgy subarea contours.

Therefore, there are two outlooks envisioned. First, imbalanced subarea sizes are accepted, smaller $N_p$ are preferred, and it is benefitted from (i) low computational complexity and (ii) naturally resulting coarse piecewise-affine subarea contours as, e.g., Fig. \ref{fig_Ex6_K6_davgdiv10} and Fig. \ref{fig_explain_3} demonstrate.

Second and alternatively, more balanced subarea sizes are preferred at the cost of high computational complexity as demonstrated in Fig. \ref{fig_Ex6_K13_davgdiv10}. Here, as an avenue of future work, a smoothing step may be added to smooth the jagged subarea contours. Two comments can be made. First, a smoothing step will inevitable introduce at least one additional hyperparameter, which is expected to be difficult to tune or generalize over different input data area scales and inter-point distances. Second, the most suitable smoothing technique appears to be to fit piecewise-affine lines to the subarea contours. Polynomial fitting appears much less suitable since it would generate convexity to some subareas but conversely then also concavities to neighboring subareas, when coinciding boundary segments between neighboring subareas are maintained. The latter subareas would complicate subsequent coverage path planning for nonholonomic vehicles.

\subsection{The dual for distributed area coverage applications\label{subsec_dual}}

As the literature review and this study revealed, area partitioning for arbitrarily shaped input data is a difficult problem. This holds in particular when resulting edgy subareas are further meant to be covered by nonholonomic vehicles with, for example, limited turning radii.

Based on this observation for distribued area coverage applications the \emph{dual problem} may be preferred. Thus, instead of focusing on area partitioning before assigning resulting subareas to robots for coverage, the dual problem instead focuses on (i) global area coverage path planning for a given work area, before then (ii) distributing sections of this path to robots. Such an approach is presented in \cite{plessen2026distributed}.

\section{Conclusion\label{sec_conclusion}}

A method for the problem of partitioning an area into K subareas subject to coinciding boundary segments of  neighboring subareas was presented. The given area was assumed to be non-convex and may additionally include obstacle areas. The proposed solution is composed of three steps, inlucing a splitting step, merging and the solution of traveling salesman problems. 

The hierarchical approach, and in particular the splitting step involving repeated bisection, permits to ensure conciding   boundary segments of neighboring subareas. 

Numerical experiments on real-world outdoors areas data revealed a tradeoff between low computational complexity but imbalanced subarea sizes, mainly as a function of the hyperparameter determining the number of repeated bisections.

\bibliography{mybibfile.bib}
\nocite{*}







\end{document}